\documentclass[preprint,aps,prd,superscriptaddress,nofootinbib]{revtex4}
\usepackage{graphicx,tabularx}
\usepackage{amsmath, amssymb, amsthm, amsfonts}
\usepackage[dvipsnames]{xcolor}
\usepackage{color}
\usepackage{multirow}
\usepackage{soul}
\usepackage{comment}
\usepackage{slashed}
\usepackage[yyyymmdd,hhmmss]{datetime}
\usepackage[backgroundcolor=white, bordercolor=white, textsize=tiny]{todonotes}
\usepackage{subfigure}

\newcommand{\pe}{\mathbf{p}_{e^-}}

\begin{document}
\title{Investigating the $Z^\prime$ gauge boson at the future lepton colliders}

\author{Xinyue Yin}
\affiliation{School of Physics and Technology, University of Jinan, Jinan Shandong 250022,  China}
\author{Honglei Li}\email{sps_lihl@ujn.edu.cn}
\affiliation{School of Physics and Technology, University of Jinan, Jinan Shandong 250022,  China}
\author{Yi Jin}
\affiliation{School of Physics and Technology, University of Jinan, Jinan Shandong 250022,  China}
\affiliation{Guangxi Key Laboratory of Nuclear Physics and Nuclear Technology, Guangxi Normal University, Guilin Guangxi 541004,  China}
\author{Zhilong Han}
\affiliation{School of Physics and Technology, University of Jinan, Jinan Shandong 250022,  China}
\author{Zongyang Lu}
\affiliation{School of Physics and Technology, University of Jinan, Jinan Shandong 250022,  China}

\begin{abstract}
$Z^\prime$ boson as a new gauge boson has been proposed in many new physics models. The interactions of $Z^\prime$ coupling to fermions are detailed studied at the large hadron collider. A $Z^\prime$ with the mass of a few TeV has been excluded in some special models. The future lepton colliders will focus on the studies of Higgs physics which provide the advantage to investigate the interactions of Higgs boson with the new gauge bosons. We investigate the $Z^\prime ZH$ interaction via the process of $e^+e^- \to Z^\prime/Z \to ZH \to l^+l^- b \bar{b}$. The angular distribution of the final leptons decaying from the $Z$-boson is related to the mixing of $Z^\prime$-$Z$ and the mass of $Z^\prime$. The forward-backward asymmetry has been proposed as an observable to investigate the  $Z^\prime$-$Z$ mixing.
The angular distributions change significantly with some special beam polarization comparing to the unpolarized condition.
\end{abstract}

\maketitle
\thispagestyle{empty}
	\setcounter{page}{1}
\section{Introduction}
With the running of the large hadron collider (LHC), the exploration of the heavy resonance, such as additional Higgs boson, neutral gauge boson ($Z^\prime$) and charged ones ($W^{\prime \pm}$), has been reached Tera-eV era currently. The Standard Model (SM) Higgs particle has been discovered in 2012 at LHC~\cite{ATLAS:2012yve, CMS:2012qbp}, which confirms the prediction of the standard model for fundamental particles and the role of Higgs in the electroweak symmetry breaking.  Further precision measurements on the Higgs particles have been proposed at the future Higgs factories, including the Circular Electron Positron Collider (CEPC)~\cite{CEPCStudyGroup:2018ghi}, the electron-positron stage of the Future Circular Collider (FCC-ee)~\cite{FCC:2018byv,FCC:2018evy}, and the International Linear Collider (ILC)~\cite{Bambade:2019fyw}. It is attractive to study the new physics effects  in  Higgs production  at the future Higgs factories.

A simple extension of the SM can be possible by adding an additional $U(1)$ group which may arise in models derived from grand unified theories (GUT). Additional $U(1)$ groups can also arise from higher dimensional constructions like string compactifications. In many models of GUT symmetry breaking, $U(1)$ groups survive at relatively low energies, leading to corresponding neutral gauge bosons, commonly referred to as $Z^\prime$ bosons~\cite{Langacker:2008yv}. Such $Z^\prime$ bosons typically couple to SM fermions via the electroweak interaction, and can be observed at hadron colliders as narrow resonances. This extra gauge boson $Z^\prime$ has been searched for many years at the Tevatron, LEP and LHC~\cite{CDF:2008xbz, D0:2010kuq, Feldman:2006ce, Chiappetta:1996km, Barger:1996kr, CMS:2019buh, ATLAS:2017fih}.
Another interesting motivation is that the  $Z^\prime$ boson as a mediator or candidate for the dark matter sectors, which is out of the reach of this paper.

There are many experimental searches for the extra gauge boson $Z^\prime$ and the strong constraints on the mass of $Z^\prime$.
The primary studied mode for a $Z^\prime$ at a hadron collider is the Drell-Yan production of a dilepton resonance $pp$ $(p\bar{p}) \to Z^\prime \to l^+ l^-$, where $l = e$ or $\mu$. Other channels, such as $Z^\prime \to jj$ where $j$ is jet, $Z^\prime \to t\bar{t}$, $Z^\prime \to e \mu$, or $Z^\prime \to \tau^+  \tau^-$, are also possible.
The forward-backward asymmetry for $pp$ $(p\bar{p})  \to l^+ l^-$ due to $\gamma$, $Z$, $Z^\prime$ interference below the $Z^\prime$ peak is also important. Nowadays, the collider phenomenology of the $Z^\prime$ boson has been extensively studied (see, for example~\cite{Rizzo:1985kn, Nandi:1986rg, Baer:1987eb, Barger:1987xw, Gunion:1987jd, Hewett:1988xc, Feldman:2006wb, Rizzo:2006nw, Lee:2008cn, Barger:2009xg, Cao:2016uur, Gulov:2018zij}).
Phenomenological implications of the $Z^\prime$ boson in an additional $U(1)$ extended model are enormous. They can be explored in fermion pair production processes in $e^-e^+$ collisions~\cite{Funatsu:2020haj,Gao:2010zzg,Das:2021esm}.
With the accumulated events increasing on the Higgs particle, the interaction of $Z^\prime$ and Higgs boson draws ones attention. An important process to study the $Z^\prime$ boson is an associated production of the SM Higgs boson ($H$) with the SM $Z$ boson, such as $p p \to Z^\prime \to ZH$ at the LHC \cite{ Li:2013ava, Das:2020rsr, ATLAS:2017xel, CMS:2021fyk} and $e^+ e^- \to Z^\prime \to ZH$ at future $e^+ e^-$ linear colliders~\cite{Gutierrez-Rodriguez:2015qka}.

To study the properties of $Z^\prime$, including one additional $U(1)_X$ extension of the SM,  we investigate the $Z^\prime$ signal via the process of $e^+e^- \to Z^\prime/Z \to ZH$ at the electron-positron collision in this paper. The cross section has been calculated including the $Z^\prime$-$Z$  mixing effects.  As the leptonic decay of $Z$ boson can be a good trigger for the signal process,  we show the final lepton angular distribution with the subsequent decay of $Z \to l^+ l^-$ and $H \to b \bar{b}$. The correlation of the final lepton angular distribution with the $Z^\prime ZH$ interactions has been investigated at the unpolarized/polarized electron-positron collision.

This paper is organized as follows.  The theoretical framework is listed in details in Section~\ref{Sec:model and exp} and the current experimental status on $Z^\prime$ has been summarized as well.  In Section~\ref{sec:ZptoZH}, we investigate the $ZH$ production via $Z^\prime/Z$ mediators and give the final leptons' angular distributions with various parameter sets. A forward-backward asymmetry has been proposed to be an observable for $Z^\prime$-$Z$ mixing. Finally, the summary and discussion are given.

\section{Theoretical framework and experimental status}{\label{Sec:model and exp}}
The gauge group structure of the standard model (SM), $SU(3)_C \times SU(2)_L \times U(1)_Y $, can be extended with an additional $U(1)_X$ group. We follow the notations in reference~\cite{Wells:2008xg} to display the interactions and mass relations. The new gauge sector $U(1)_X$ is only coupled with the standard model by kinetic mixing with  hypercharge gauge boson $B_\mu$.
The kinetic energy terms of the $U(1)_X$ gauge group are
\begin{align}
  \begin{split}
    \mathcal{L}_K
    &=- \frac{1}{4} \hat{X}_{\mu\nu} \hat{X}^{\mu\nu} + \frac{\chi}{2} \hat{X}_{\mu\nu} \hat{B}^{\mu\nu}
  \end{split} ,
\end{align}
where  $\chi \ll 1$ is helpful to keep precision electroweak predictions consistent with experimental measurements. And the Lagrangian for the gauge sector is given by
\begin{align}
  \begin{split}
    \mathcal{L}_G
    &=- \frac{1}{4} \hat{B}_{\mu\nu} \hat{B}^{\mu\nu} - \frac{1}{4} \hat{W}_{\mu\nu}^a \hat{W}^{a\mu\nu} - \frac{1}{4} \hat{X}_{\mu\nu} \hat{X}^{\mu\nu} + \frac{\chi}{2} \hat{X}_{\mu\nu} \hat{B}^{\mu\nu} ,
  \end{split}
\end{align}
where $W_{\mu\nu}^a$, $B_{\mu\nu}$, $X_{\mu\nu}$ are the field strength tensors for $SU(2)_L$, $U(1)_Y$, $U(1)_X$, respectively.
We can diagonalize the 3$\times$3 neutral gauge boson mass matrix, and write the mass eigenstates as
\begin{align}
  \begin{split}
    \begin{pmatrix}
        B & \\ W^3 & \\ X
       \end{pmatrix}
    &=\begin{pmatrix}
        \cos \theta_W & -\sin \theta_W \cos \alpha & \sin \theta_W \sin \alpha\\
        \sin  \theta_W & \cos \theta_W \cos \alpha & -\cos \theta_W \sin \alpha\\
        0 & \sin \alpha & \cos \alpha \\
       \end{pmatrix}
       \begin{pmatrix}
       A & \\ Z & \\ Z^\prime
       \end{pmatrix}
  \end{split} ,
\end{align}
where the usual weak mixing angle and the new gauge boson mixing angle are
\begin{align}
  \begin{split}
    \sin \theta_W
    &=\frac{g^\prime}{\sqrt{g^2 + g^{\prime2}}} \ ;\ \ \tan 2\alpha = \frac{-2 \eta \sin \theta_W}{1- \eta^2 \sin^2 \theta_W - \Delta z} \
  \end{split},
\end{align}
with $\Delta z$ = $\frac{m_X^2}{m_{Z_0}^2}$, 
$m_{Z_0}^2$ = $\frac{(g^2 + g^{\prime2})v^2}{4}$, $\eta = \frac{\chi}{\sqrt{1-\chi^2}}$. $m_{Z_0}$ and $m_X$ are masses before mixing. The two heavier boson mass eigenvalues are
\begin{align}
  \begin{split}
    m_{Z,Z^\prime}^2
    &=\frac{m_{Z_0}^2}{2} \left[ \left( 1 + \eta^2  \sin^2 \theta_W + \Delta Z \right) \pm \sqrt{\left(1 - \eta^2  \sin^2 \theta_W - \Delta Z \right)^2 + 4 \eta^2 \sin^2 \theta_W } \right]
  \end{split} .
\end{align}
With the assumption of $ \eta \ll 1$, we can take the mass eigenvalues as $m_Z \approx m_{Z_0} = 91.19$GeV and $m_{Z^\prime} \approx m_X$.
Therefore, the $Z$ and $Z^\prime$ coupling to SM fermions are
\begin{align}
  \begin{split}
    \bar{\psi} \psi Z
    &=\frac{ig}{\cos \theta_W} \left[ \cos \alpha \left( 1 - \eta \sin \theta_W \tan \alpha  \right) \right] \left[T_L^3 - \frac{\left(1 - \eta \tan \alpha / \sin  \theta_W \right)}{\left(1 - \eta  \sin \theta_W \tan \alpha \right)} \sin^2 \theta_W Q \right]
  \end{split} ,
\end{align}
\begin{align}
  \begin{split}
    \bar{\psi} \psi Z^\prime
    &=\frac{-ig}{\cos \theta_W} \left[ \cos \alpha \left(\tan \alpha + \eta \sin \theta_W \right) \right] \left[T_L^3 - \frac{\left(\tan \alpha + \eta / \sin  \theta_W \right)}{\left(\tan \alpha + \eta \sin \theta_W \right)} \sin^2 \theta_W Q \right]
  \end{split} ,
\end{align}
and $Z^\prime ZH$ interaction can be  written as
\begin{align}
  \begin{split}
    Z^\prime ZH
    &= 2i \frac{m_{Z_0}^2}{v} \left(-\cos \alpha + \eta \sin \theta_W
    \sin \alpha \right) \left( \sin \alpha + \eta \sin \theta_W \cos \alpha
    \right).
  \end{split}
\end{align}

The above is a very common model including $Z^\prime$ boson. There are also such kinds of models which differ from the coupling strength, e.g., $Z^\prime_{SSM}$, $Z^\prime_{\psi}$, etc., which have been studied extensively at high energy colliders.

The processes of $pp \to Z^\prime \to e^+ e^-$ and $pp \to Z^\prime \to \mu^+ \mu^-$  are studied by the D0~\cite{D0:2010kuq} and CDF collaborations~\cite{CDF:2008jev}, respectively, and lower limit on the mass of the  Sequential Standard Model (SSM) $Z^\prime$ boson of 1023 GeV is given by D0. The CDF excludes the mass region of 100 GeV $ < m_{Z^\prime} < $ 982 GeV for a $Z^\prime_{\eta}$ boson in the E6 model.

For the SSM $Z^\prime$ bosons decaying to a pair of $\tau$-leptons with $m_{Z^\prime}$ $\textless$ 2.42 TeV is excluded at 95$\%$ confidence level by the ATLAS experiment, while $m_{Z^\prime}$ $\textless$ 2.25 TeV is excluded for the non-universal $G(221)$ model that exhibits enhanced couplings to third-generation fermions \cite{ATLAS:2017eiz}.
And searches are performed for high-mass resonances in the dijet invariant mass spectrum. $Z^\prime$ gauge bosons are excluded in the SSM with $Z^\prime \to b\bar{b}$ for masses up to 2.0 TeV, and excluded in the leptophobic model with SM-value couplings to quarks for masses up to 2.1 TeV~\cite{ATLAS:2018tfk}.

It shows that the results imply a lower limit of 4.5 (5.1) TeV on $m_{Z^\prime}$ for the E6-motivated $Z^\prime_\psi$ ($Z^\prime_{SSM}$) boson in the dilepton channel~\cite{ATLAS:2019erb}.
Upper limits are set on the production cross section times branching fraction for the $Z^\prime$ boson in the top color-assisted-technicolor model, resulting in the exclusion of $Z^\prime$ masses up to 3.9 TeV and 4.7 TeV for decay widths of 1$\% $ and 3$\%$ , respectively \cite{ATLAS:2020lks}.
The heavy $Z^\prime$ gauge  bosons  decay  into  $e\mu$ final  states in proton-proton collisions are excluded for masses up to 4.4 TeV by the CMS experiment~\cite{CMS:2018hnz}.  By analyzing data of proton-proton collision collected by the CMS experiment from 2016 to 2018, the limit of $Z^\prime$ mass is obtained for the combination of the dielectron and dimuon channels. The limits are 5.15 TeV for the $Z^\prime_{SSM}$ and 4.56 TeV for the $Z^\prime_\psi~$\cite{CMS:2021ctt}.

The studies on the $Z^\prime$ boson also conducted at the lepton colliders especially in the small mass regions. The experiments of ALEPH and OPAL at the LEP limited the mass of the $Z^\prime$ boson with the process of $e^+e^- \to ff$ where $f$ is a $\tau$ lepton or $b$ quark. The lower limits of $Z^\prime$ from $\tau$-pair production are 365 GeV at ALEPH and 355 GeV at OPAL, and those from $b\bar{b}$ production are 523 GeV at ALEPH and 375 GeV at OPAL~\cite{Lynch:2000md}. With the proposal of the future lepton colliders, the mass region and the coupling strength of the $Z^\prime$ boson can be extended extensively, especially in the process  of Higgs boson production.

\section{ $Z^\prime ZH$ interaction at the lepton colliders}
\label{sec:ZptoZH}
\subsection{$e^+e^- \to Z^{\prime}/Z \to ZH$}

\begin{figure}[h]
  \centering
  \subfigure[ ]{\hspace{-5mm}
   \includegraphics[width=0.5\textwidth]{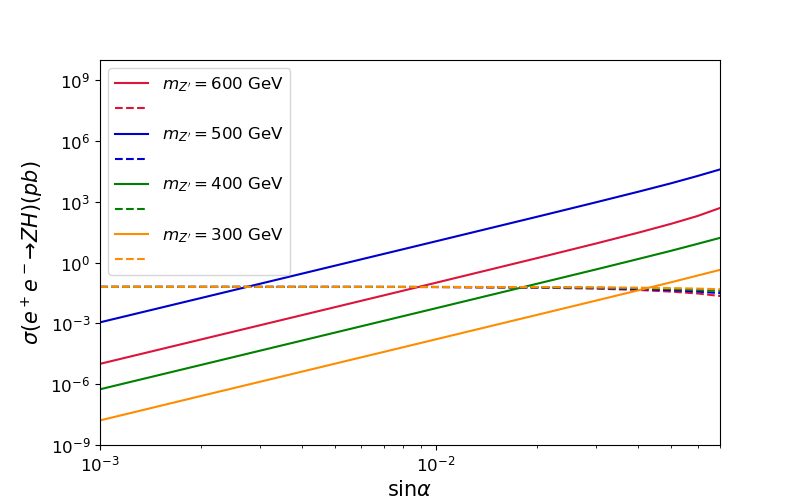}
   }
   \subfigure[ ]{\hspace{-5mm}
   \includegraphics[width=0.5\textwidth]{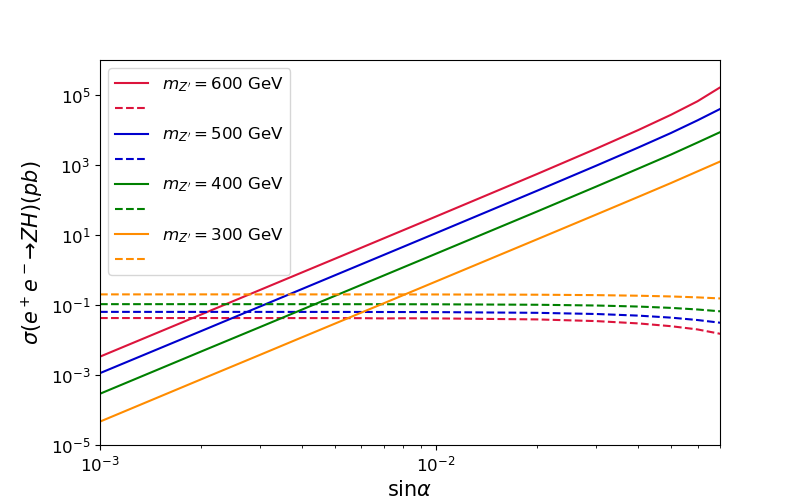}
   }
  \caption{
     The cross section of $e^+e^- \to Z \to ZH$ (dashed) and
     $e^+e^- \to Z^\prime \to ZH$ (solid) versus $\sin\alpha$ with $\sqrt{s} = 500$ GeV for the left panel. The right panel is the same plot as the left one with $\sqrt{s} = m_{Z^\prime}$.
  }
\label{fig:section1}
\end{figure}

\begin{figure}[htb]
  \centering
  \includegraphics[width=0.6\textwidth]{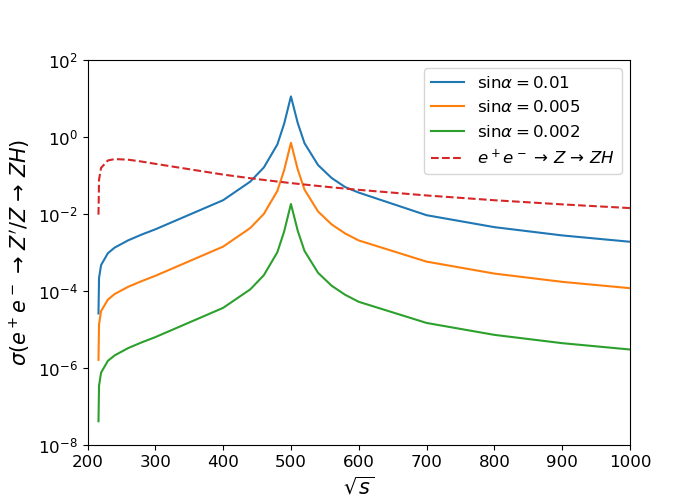}

  \caption{
    The cross section of processes $e^+e^- \to Z^\prime/Z \to ZH$ with $m_{Z^\prime}=500$ GeV.
  }
  \label{fig:resonance}
\end{figure}

In this section, we consider the process of $e^+e^- \to Z^{\prime}/Z \to ZH$ at the future lepton colliders. The cross sections are displayed in Fig.\ref{fig:section1} as a function of $\sin\alpha$.
The mass of $Z^\prime$ is 300, 400, 500 and 600 GeV respectively. The cross section for $m_{Z^\prime}= 500$ GeV is the largest one due to the resonance enhanced effects with the center-of-mass-system (C.M.S.) energy $\sqrt{s}=500$ GeV. We can notice that the cross section is 11.33 pb with $m_{Z^\prime} = 500$ GeV when $\sin\alpha$ is 0.01. The cross section of $e^+e^- \to Z \to ZH$ is closed to 0.1 pb with a slight decrease when $\sin\alpha$ get large, that is because the interaction of $Z$ and Higgs boson has a slight dependence on the mixing angle.
The resonance threshold effects are obvious with the $\sqrt{s} = m_{Z^\prime}$  from the lines in the right panel comparing with the left ones.
In Fig.\ref{fig:resonance},  we show the cross section distribution with various C.M.S. energies for $m_{Z^\prime} = 500$ GeV. With the increase of $\sin\alpha$, the cross section of  $e^+e^- \to Z^\prime \to ZH$ gets large. If the C.M.S. energy is set around the value of $m_{Z^\prime}$, the cross section of  $e^+e^- \to Z^\prime \to ZH$ will be three or four orders larger than that without resonance effects.
\subsection{Angular distribution of the final leptons}
With the cascaded decay of $H \to b \bar{b}$ and $Z \to l^+ l^-$,  we choose the final state of $b\bar{b} l^+l^-$ as the final signature at the collider. Following the discussion in the reference of~\cite{Li:2013ava}, we adopt the angular distribution of the final leptons to investigate the $Z^\prime ZH$ interaction. The angle can be expressed from the following formula,
\begin{align}
  \begin{split}
    \cos\theta
    &= \frac{ \mathbf{p}^*_{l^-} \cdot \pe }{ \lvert \mathbf{p}^*_{l^-} \rvert \cdot \lvert \pe \rvert}\ ,
  \end{split}
\end{align}
where $\mathbf{p}^*_{l^-}$ is the three-momentum of the negative charged lepton in the $Z$ boson rest frame and $\pe$ is the three-momentum of the electron in the $e^+e^-$ rest frame.

\begin{figure}[b]
  \centering
  \subfigure[ $\sqrt{s}$ = 500 GeV]{\hspace{-8mm}
   \includegraphics[width=0.5\textwidth]{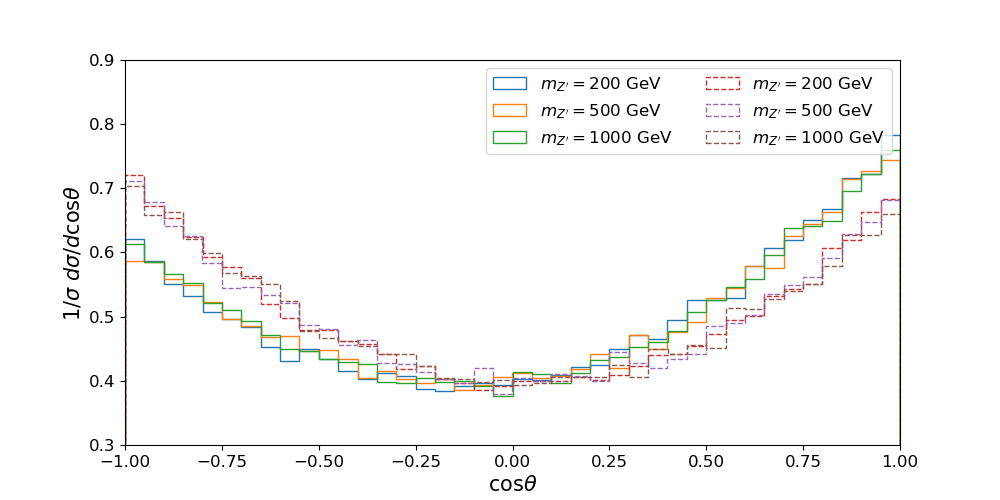}

   \label{fig:etas=500}
   }
   \subfigure[$\sqrt{s}$ = 1000 GeV]{\hspace{-8mm}
   \includegraphics[width=0.5\textwidth]{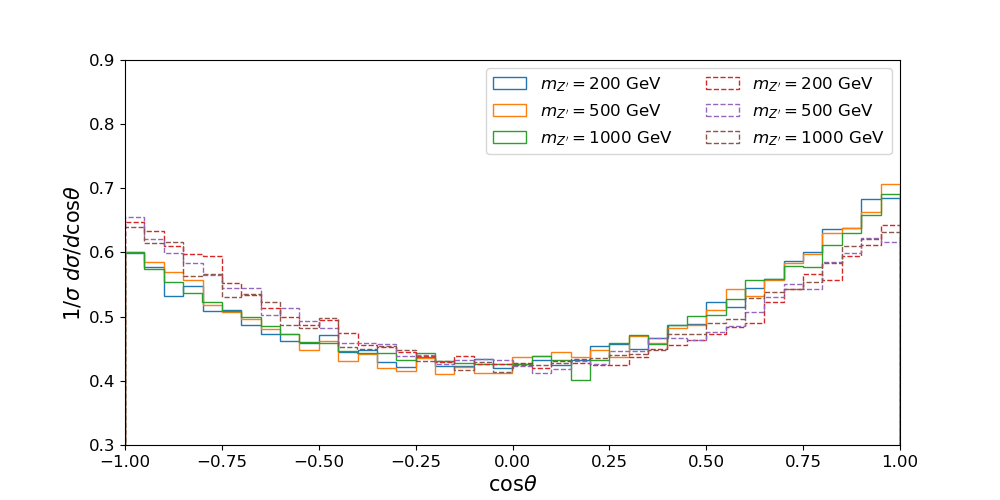}
   \label{fig:etas=1000}
   }
  \caption{
     The angular distribution for the charge lepton $l^-$ with $\eta=0.1$. The C.M.S. energy is 500 GeV (a) and 1000 GeV (b). The solid lines are the distributions for process $e^+e^- \to Z^\prime \to ZH \to l^+l^- b \bar{b}$ and the dashed lines are for $e^+e^- \to Z^\prime /Z\to ZH \to l^+l^- b \bar{b}$  with $m_{Z^\prime}$  taken as 200 GeV, 500 GeV and 1000 GeV  respectively.
  }
\label{fig:etas}
\end{figure}

Fig.\ref{fig:etas} shows the angular distribution $1/\sigma d\sigma/d\cos\theta$ for the charged lepton $l^-$ in the process of $e^+e^- \to Z^\prime /Z\to ZH \to l^+l^- b \bar{b}$ with  $\eta=0.1$. The distributions with pure $Z^\prime$ contributions are well separated from those with  both $Z^\prime$ and $Z$-boson mediators effects in Fig.\ref{fig:etas} (a) with $\sqrt{s}=500$ GeV. The angular distributions are not very sensitive to the mass of $Z^\prime$ bosons. Comparing the distributions with different C.M.S. energies from Fig.\ref{fig:etas=500} and Fig.\ref{fig:etas=1000}, the discrepancy is reduced with the increase of the collision energy.
A forward-backward asymmetry can be defined as
\begin{align}
  \begin{split}
    A_{FB}
    &= \frac{ \sigma(\cos \theta \ge 0 ) - \sigma(\cos \theta < 0) }{  \sigma(\cos \theta \ge 0) + \sigma(\cos \theta < 0)} \ .
  \end{split}
\end{align}
Corresponding to the studies in Fig.\ref{fig:etas}, the asymmetries are listed in Table \ref{tab:etas}. The forward-backward asymmetry is 0.0726 with $\sqrt{s} = m_{Z^\prime} = 500$ GeV for the pure $Z^\prime$ contribution, while it changes to $-0.0136$ after considering the $Z$-boson contributions.

\begin{table}[!h]
\centering
\begin{tabular}{|c|c|c|c|c|c|c|}
\hline
\multirow{2}{*}{}&
\multicolumn{3}{c|}{$\sqrt{s}=500$ GeV} &
\multicolumn{3}{c|}{\multirow{1}{*}{$\sqrt{s}=$ 1000 GeV}}\\
\cline{1-7}
  $m_{Z^\prime}$ &  200 GeV  &  500 GeV  &  1000 GeV  &  200 GeV  &  500 GeV  &  1000 GeV    \\
\hline
$ Z^\prime $ & $ 8.51 \times 10^{-2}$& $7.26\times 10^{-2}$ & $7.14\times 10^{-2} $& $4.94\times 10^{-2}$& $4.02\times 10^{-2}$& $3.83\times 10^{-2}$ \\
\hline
$Z/Z^\prime$ & $-2.09\times 10^{-2}$& $-1.36\times 10^{-2}$ & $-2.47\times 10^{-2}$ & $-1.05\times 10^{-2}$ & $-6.69\times 10^{-3}$ & $-4.98\times 10^{-3}$\\
\hline
\end{tabular}
  \caption{
     The forward-backward asymmetry for process of $e^+ e^- \to Z^\prime \to ZH \to l^+l^-b\bar{b}$ and $e^+ e^- \to Z^\prime/Z \to ZH \to l^+l^-b\bar{b}$ with $\eta=0.1$.
  }
\label{tab:etas}
\end{table}
\begin{figure}[h]
  \centering
  \includegraphics[width=1\textwidth]{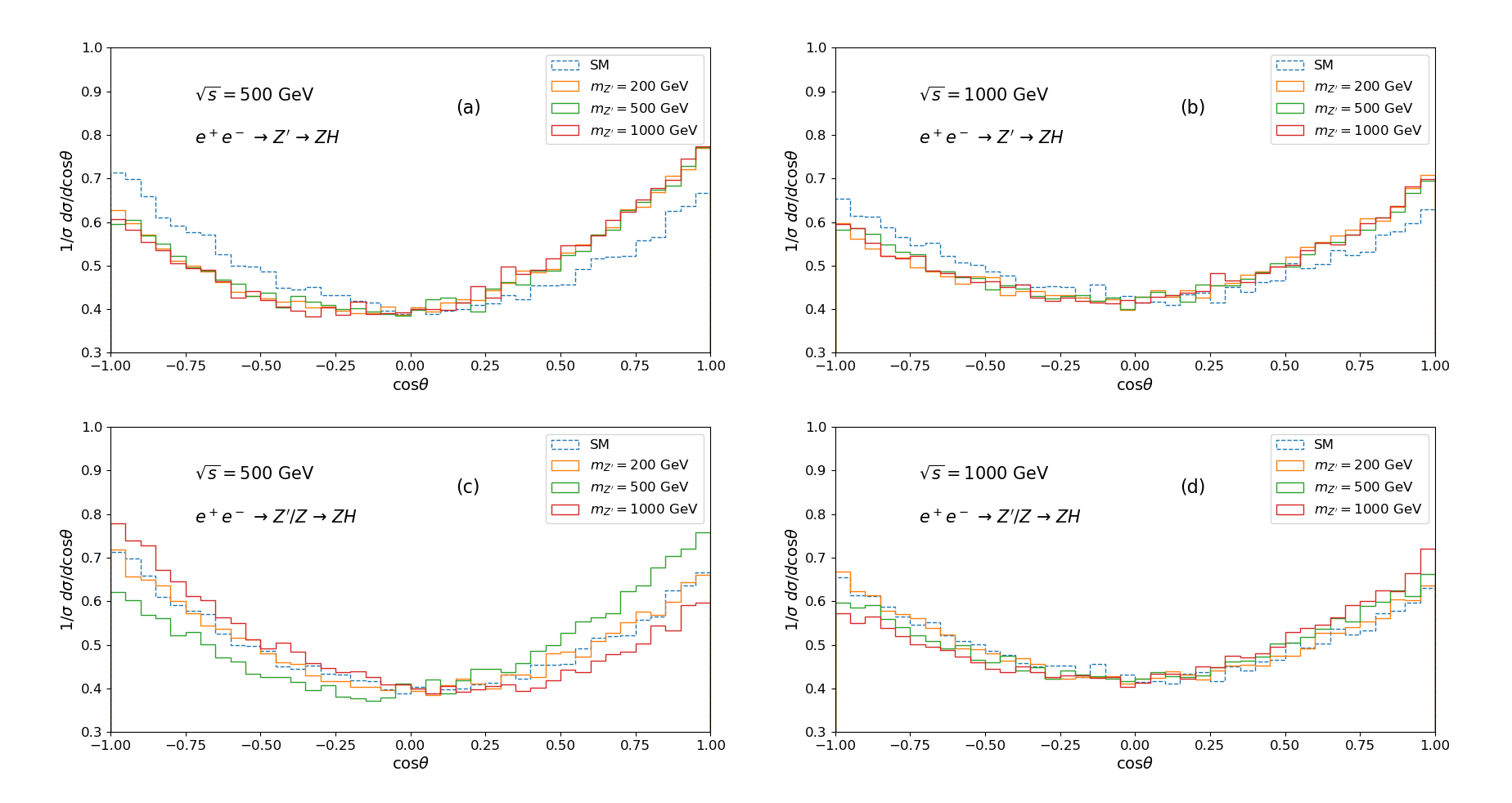}

  \caption{
     The angular distribution for the charge lepton $l^-$ with $\sin\alpha= 0.01$. The solid lines are the distributions for process $e^+e^- \to Z^\prime \to ZH \to l^+l^- b \bar{b}$ or $e^+e^- \to Z^\prime/Z \to ZH \to l^+l^- b \bar{b}$ with $m_{Z^\prime}$  taken as 200 GeV, 500 GeV and 1000 GeV  respectively and the dashed lines are the SM process $e^+e^- \to Z \to ZH \to l^+l^- b \bar{b}$.
  }
\label{fig:mzp}
\end{figure}

The angular distributions of the charge lepton $l^-$ are showed in Fig.\ref{fig:mzp} with $\sin\alpha = 0.01$. The distributions with $\cos\theta$ $\textless$ 0 are lower than those with $\cos\theta \ge 0$  for $e^+ e^- \to Z^\prime \to ZH \to l^+l^-b\bar{b}$ in Fig.\ref{fig:mzp} (a) and (b). And the distributions are separated obviously with various $m_{Z^\prime}$ comparing with Fig.\ref{fig:mzp} (a) and Fig.\ref{fig:mzp} (c), where the contribution from $Z$-boson and $Z^\prime$-$Z$ interference are included in Fig.\ref{fig:mzp} (c). With increase of the C.M.S. energy as Fig.\ref{fig:mzp} (d), the discrepancies are lessened but the asymmetries can be found with the resonance effects when $m_{Z^\prime} = \sqrt{s} = 1000$ GeV for $e^+e^- \to Z^\prime/Z \to ZH \to l^+l^- b \bar{b}$. The corresponding forward-backward asymmetries are listed in  Table \ref{tab:mzp}.

\begin{table}[!h]
\centering
\begin{tabular}{|c|c|c|c|c|c|c|}
\hline
\multirow{2}{*}{}&
\multicolumn{3}{c|}{$\sqrt{s}=500$ GeV} &
\multicolumn{3}{c|}{\multirow{1}{*}{$\sqrt{s}=$ 1000 GeV}}\\
\cline{1-7}
  $m_{Z^\prime}$ &  200 GeV  &  500 GeV  &  1000 GeV  &  200 GeV  &  500 GeV  &  1000 GeV    \\
\hline
$ Z^\prime $ & $ 8.24\times 10^{-2}$& $7.29\times 10^{-2}$ & $9.41\times 10^{-2} $& $4.93\times 10^{-2}$& $4.22\times 10^{-2}$& $4.82\times 10^{-2}$ \\
\hline
$Z/Z^\prime$ & $-2.17\times 10^{-2}$& $7.77\times 10^{-2}$ & $-8.72\times 10^{-2}$ & $-1.20\times 10^{-2}$ & $3.10\times 10^{-2}$ & $5.17\times 10^{-2}$\\
\hline
\end{tabular}
  \caption{
     The forward-backward asymmetry for process of $e^+ e^- \to Z^\prime \to ZH \to l^+l^-b\bar{b}$ and $e^+ e^- \to Z^\prime/Z \to ZH \to l^+l^-b\bar{b}$ with $\sin\alpha=0.01$.
  }
\label{tab:mzp}
\end{table}

\begin{figure}[h]
  \centering
  \includegraphics[width=1\textwidth]{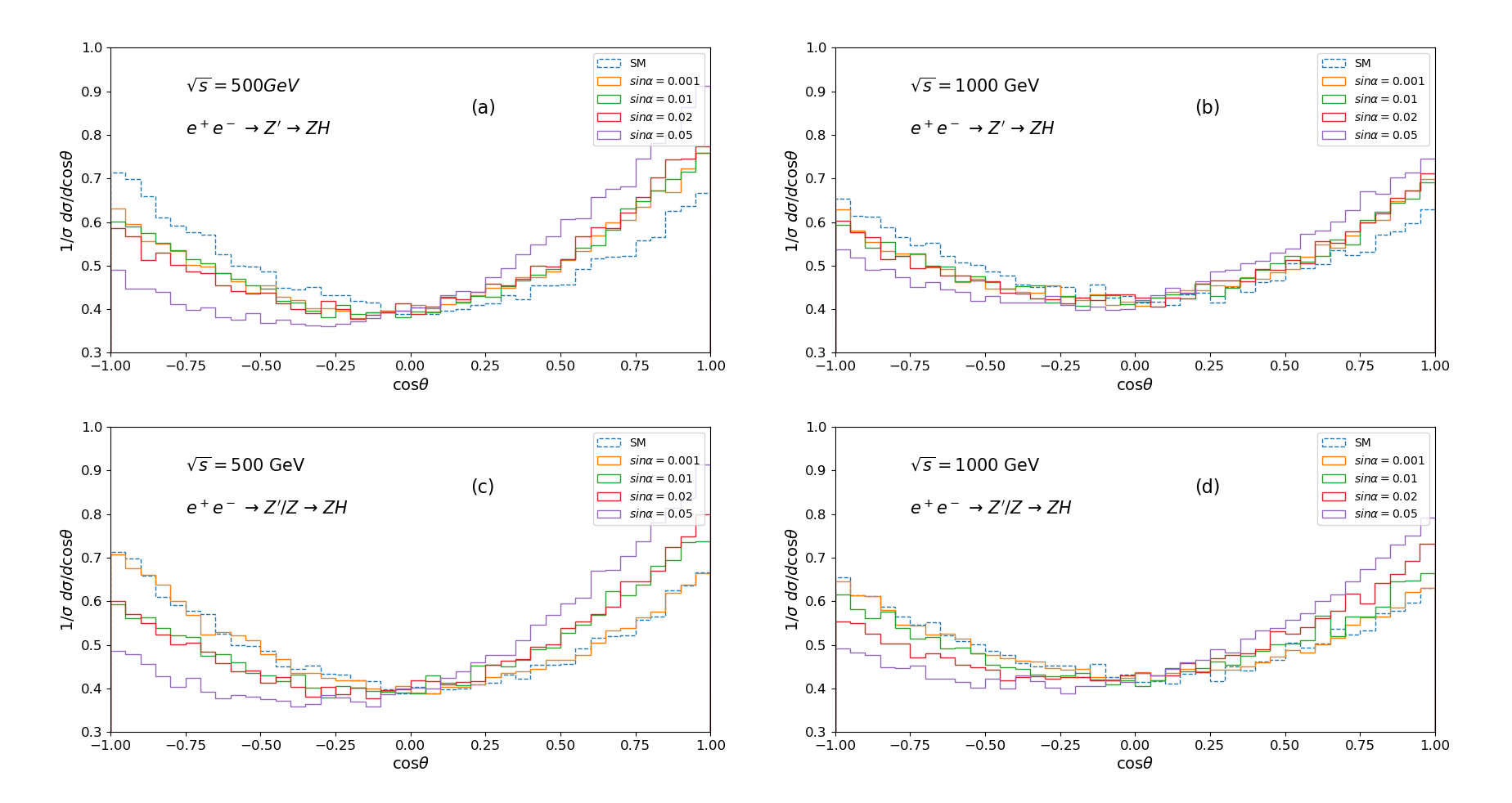}

  \caption{
       The angular distribution for the charge lepton $l^-$ with $m_{Z^\prime}= 500$ GeV. The solid lines are the distributions for process $e^+e^- \to Z^\prime \to ZH \to l^+l^- b \bar{b}$ or $e^+e^- \to Z^\prime/Z \to ZH \to l^+l^- b \bar{b}$ with $\sin\alpha$ are taken as 0.001, 0.01, 0.02 and 0.05 respectively and the dashed lines are for SM process $e^+e^- \to Z \to ZH \to l^+l^- b \bar{b}$.
  }
\label{fig:sina}
\end{figure}

In Fig.\ref{fig:sina} we show the angular distributions of $e^+e^- \to Z^\prime \to ZH \to l^+l^-b\bar{b}$ and $e^+e^- \to Z^\prime/Z \to ZH \to l^+l^-b\bar{b}$ with different  $Z^\prime$-$Z$ mixing angles. With the increase of the mixing angle, the asymmetry becomes large with $\sqrt{s}=500$ GeV. The asymmetry for  $e^+e^- \to Z^\prime \to ZH$  is close to the SM process of $e^+e^- \to Z \to ZH$ with the small mixing angle displayed in Fig.\ref{fig:sina} (a), and the same performance are showed within $Z$-boson effects from Fig.\ref{fig:sina} (c). The asymmetry of the process $e^+ e^- \to Z^\prime/Z \to ZH \to l^+l^-b\bar{b}$ with $\sin\alpha=0.001$ is hardly distinguished from the asymmetry of pure $Z$-boson contribution. With the increase of the collision energy, the distributions show the same tendency in Fig.\ref{fig:sina} (b) and (d) except that the discrepancies become small comparing with Fig.\ref{fig:sina} (a) and (b).
The detailed values are listed in Table.\ref{tab:sina} with different collision energy and $\sin\alpha$. The asymmetry is 0.0722 (0.213) with $\sin\alpha=0.001$ ($\sin\alpha=0.05$) for $e^+e^- \to Z^\prime \to ZH \to l^+l^-b\bar{b}$ process with $\sqrt{s}=500$ GeV. Both including the $Z$ and $Z^\prime$ effects, the asymmetry is $-0.0223$ (0.210) with $\sin\alpha=0.001$ ($\sin\alpha=0.05$) for $e^+e^- \to Z^\prime /Z \to ZH \to l^+l^-b\bar{b}$ process with $\sqrt{s}=500$ GeV.

\begin{table}[!h]
\centering
\begin{tabular}{|c|c|c|c|c|}
\hline
\multicolumn{5}{|c|}{$\sqrt{s}=500$ GeV} \\
\cline{1-5}
  $\sin\alpha$ & $ 0.001 $ & $ 0.01 $ & $ 0.02 $ & $ 0.05 $  \\
\hline
$ Z^\prime $ & $ 7.22 \times 10^{-2}$& $7.29\times 10^{-2}$ & $9.30\times 10^{-2} $& $2.13\times 10^{-1}$\\
\hline
$Z/Z^\prime$ & $-2.23\times 10^{-2}$& $7.77\times 10^{-2}$ & $9.21\times 10^{-2}$ & $2.10\times 10^{-1}$\\
\hline
\multicolumn{5}{|c|}{$\sqrt{s}=1000$ GeV} \\
\cline{1-5}
  $\sin\alpha$ & $ 0.001 $ & $ 0.01 $ & $ 0.02 $ & $ 0.05 $  \\
\hline
$ Z^\prime $ & $ 4.18\times 10^{-2}$& $4.22\times 10^{-2}$ & $4.94\times 10^{-2} $& $1.16\times 10^{-1}$\\
\hline
$Z/Z^\prime$ & $-7.43\times 10^{-3}$& $3.10\times 10^{-2}$ & $8.29\times 10^{-2}$ & $1.39\times 10^{-1}$\\
\hline
\end{tabular}
  \caption{
     The forward-backward asymmetry for process of $e^+ e^- \to Z^\prime \to ZH \to l^+l^-b\bar{b}$ and $e^+ e^- \to Z^\prime/Z \to ZH \to l^+l^-b\bar{b}$ with $m_{Z^\prime}$ is set at 500 GeV.
  }
\label{tab:sina}
\end{table}

\subsection{Signals with beam polarization}

\begin{figure}[h]
  \centering
  \includegraphics[width=0.6\textwidth]{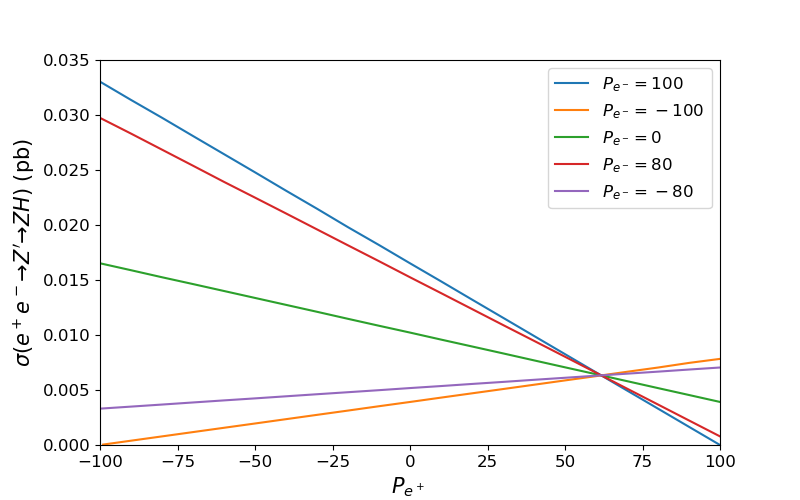}

  \caption{
  The cross section of $e^+e^- \to Z^\prime \to ZH$ with polarized electron and positron beam.
    }
  \label{fig:polcs}
\end{figure}

The lepton colliders have the advantage at the polarization of electron/positron beams and somehow it can affect the measurement on the final particles' distribution. We consider the production of $e^+e^- \to Z^\prime \to ZH$ with the beam polarization.  The cross sections for the studied process are displayed in Fig.\ref{fig:polcs} with various polarizations of  positron and a fixed value of electron. $P_{e^{\pm}}$ denotes the longitudinal polarization of $e^{\pm}$. $P_{e^{+}}=100$ corresponds to purely right-handed $e^+$.
The cross sections are decreased with the positron's polarization changed from $-100$ to 100 when the electron is right-handed, while the inverse trends are showed when electron is left-handed. There is an interesting point with $P_{e^+}\sim 62$ that all the lines are crossing together. This is due to the dependence on the polarization being cancelled at this point. Besides, the cross section gets small when the positron and the electron have the same sign polarizations.

\begin{figure}[h]
  \centering

   \subfigure[]{\hspace{-10mm}
   \includegraphics[width=0.5\textwidth]{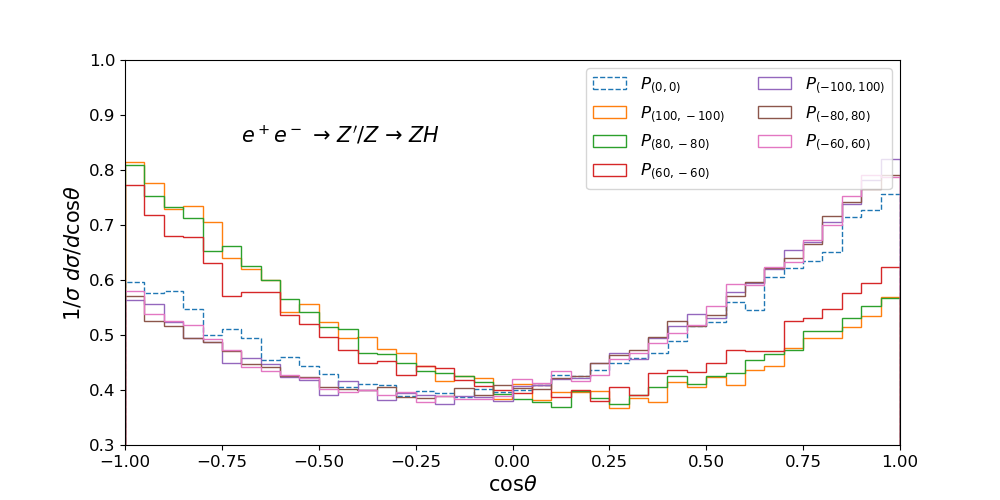}
   \label{fig:pol}
   \hspace{-10mm}
   }
   \subfigure[]{\hspace{-2mm}
   \includegraphics[width=0.5\textwidth]{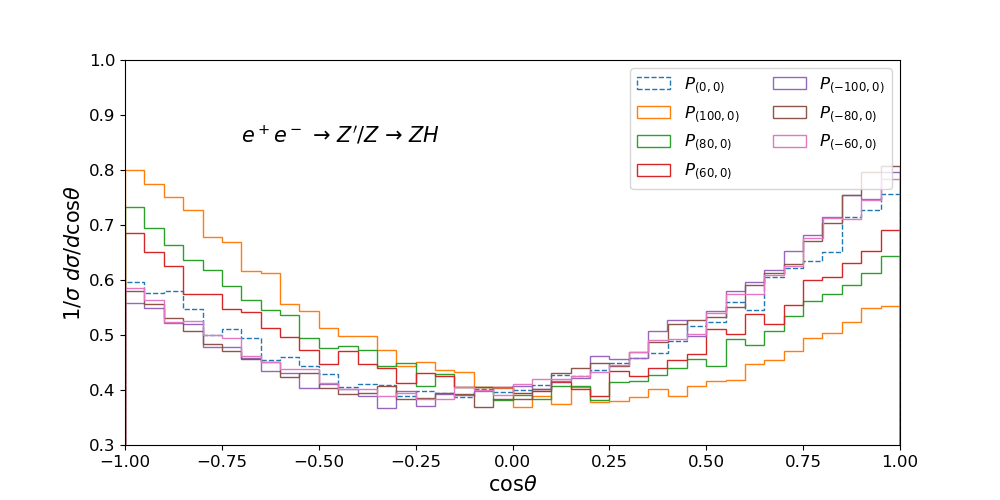}
   \label{fig:e+pol}
   }
  \caption{
  The angular distribution for process $e^+ e^- \to Z^\prime/Z \to ZH \to l^+l^-b\bar{b}$ with $m_{Z^\prime}=\sqrt{s}=500$ GeV and $\sin\alpha=0.01$. The colored lines correspond to various polarization values ($P_{e^+}$, $P_{e^-}$).
  }
\label{fig:pola}
\end{figure}

The angular distribution is displayed in Fig.\ref{fig:pola} for process $e^+ e^- \to Z^\prime/Z \to ZH \to l^+l^-b\bar{b}$ with  various beam polarizations. The parameters are set as $m_{Z^\prime} = \sqrt{s} = 500$ GeV and $\sin\alpha=0.01$. Comparing with the unpolarized condition, the distributions change obviously especially with the right-handed positron and left-handed electron collisions. Supposing the initial state with  right-handed polarized positron and unpolarized electron, the distinctions are significantly in the distributions from Fig.\ref{fig:e+pol}. The forward-backward asymmetries have been listed in Table~\ref{tab:pola}. The asymmetry can reach $-0.124$ with one hundred percents polarization of right-handed positron and left-handed electron.

\begin{table}[!h]
\centering
\begin{tabular}{|c|c|c|c|}

\cline{1-4}
  ($P_{e^+}$, $P_{e^-}$)& $A_{FB} $ &  ($P_{e^+}$, $P_{e^-}$)  & $ A_{FB} $ \\
\hline
\hline
$(100,-100) $ & $ -1.24 \times 10^{-1}$ & $(100,0)$ & $-1.23 \times 10^{-1} $ \\
\hline
 $(80,-80)$ & $ -1.10 \times 10^{-1}$ & $(80,0)$ & $-4.51 \times 10^{-2} $ \\
\hline
$(60,-60)$ & $ -6.90 \times 10^{-2}$& $(60,0)$ & $2.70 \times 10^{-3}$\\
\hline
 $(-100,100)$& $ 1.27 \times 10^{-1}$ & $(-100,0)$& $ 1.24 \times 10^{-1}$ \\
\hline
$(-80,80)$ & $1.22\times 10^{-1}$& $(-80,0)$ & $1.18 \times 10^{-1}$ \\
\hline
$(-60,60)$& $1.19\times 10^{-1} $ & $(-60,0)$& $ 1.10 \times 10^{-1}$ \\
\hline
\end{tabular}
  \caption{
     The forward-backward asymmetry for process of $e^+ e^- \to Z^\prime/Z \to ZH \to l^+l^-b\bar{b}$ with $m_{Z^\prime}=\sqrt{s}=500$ GeV and $\sin\alpha=0.01$.
  }
\label{tab:pola}
\end{table}

\section{Summary and Discussion}

The extensions of the standard model have been studied extensively at the frontier of energy and luminosity of high energy collision experiments. $Z^\prime$ boson as a new gauge boson has been proposed in many new physics models. As a proton-proton collider, the LHC has reported the searching results of $Z^\prime$ with the mass of a few TeV. The interactions for $Z^\prime$ coupling to fermions have been investigated in detail. The next generation of high energy colliders possibly focuses on the Higgs physics, i.e., Higgs-factory, which provides an opportunity to study the interaction of $Z^\prime$ coupling to Higgs boson.

In this paper we investigated the process of $e^+ e^- \to Z^\prime/Z \to ZH \to l^+l^-b\bar{b}$. The interactions and couplings are followed up the models proposed by J.D. Wells et al.. The $Z^\prime$ couplings to the standard models are related to the $Z^\prime$-$Z$ mixing.   We investigate the $e^+ e^- \to Z^\prime \to ZH$ production cross section. The cross section will be at the same order as the standard model process  $e^+ e^- \to Z \to ZH$ with the mixing angle $\sin\alpha\sim10^{-2}$. The angular distributions of the leptons decaying from the $Z$-boson rely on the mixing angle and $Z^\prime$ mass, we have investigated the distributions with the parameters variation of $0.001 < \sin\alpha < 0.1$, $200~\text{GeV} < m_{Z^\prime} < 1000$ GeV and $\sqrt{s} = 500, 1000$ GeV. The forward-backward asymmetry can reach 0.0729 for $\sin\alpha = 0.01$ and $m_{Z^\prime} = \sqrt{s} = 500$ GeV. The beam polarization effects have been investigated for the signal process $e^+ e^- \to Z^\prime/Z \to ZH \to l^+l^-b\bar{b}$. The final particles distributions change obviously with some special polarization comparing to the unpolarized condition. As the raising of the precision measurement on the Higgs boson, these studies will promote the understanding of the interactions between Higgs particle and the new physics ones.

\section*{Ackonwledgement}
This work was supported by the National Natural Science Foundation of China (NNSFC) under grant Nos.~11635009 and~11805081, Natural Science Foundation of Shandong Province under grant No.~ZR2019QA021, and the Open Project of Guangxi Key Laboratory of Nuclear Physics and Nuclear Technology, No.~NLK2021-07.

\bibliography{references}

\end{document}